\documentclass[]{spie}  %>>> use for US letter paper
\usepackage{amsmath,amsfonts,amssymb}
\usepackage{graphicx}
\usepackage[colorlinks=true, allcolors=blue]{hyperref}
\usepackage{float}

\title{Accurate uncertainty estimation in crowded fields: adaptive optics and speckle data}

\author[a]{E. Gallego-Cano}
\author[a]{R. Sch\"odel}
\author[a]{A. T. Gallego-Calvente}
\author[b]{A. M. Ghez}
\affil[a]{Instituto de Astrof\'isica de Andaluc\'ia (CSIC),
     Glorieta de la Astronom\'ia s/n, 18008 Granada, Spain}
\affil[b]{UCLA, Department of Physics and Astronomy, Los Angeles, CA 90095, USA}

\authorinfo{E. Gallego-Cano: E-mail: lgc$@$iaa.es}

\begin{document} 
\maketitle

\begin{abstract}
Optimal error estimation is key to achieve accurate photometry and astrometry. Stellar fluxes and positions in high angular resolution images are typically measured with PSF fitting routines, such as {\it StarFinder}. However, the formal uncertainties computed by these software packages tend to seriously underestimate the relevant uncertainties. We present a new approach to deal with this problem using a resampling method to obtain robust and reliable uncertainties without loss of sensitivity.
\end{abstract}


\section{INTRODUCTION}
\label{sec:intro}  % \label{} allows reference to this section

In order to obtain the photometry and astrometry of crowded stellar fields, we can use different Point Spread Function (PSF) fitting routines. In the present work, we use {\it StarFinder} \cite{diolaiti2000analysis} to analyse two different data sets: adaptive optics (AO) and speckle data. Although the program gives a reliable detection of the point sources and determination of their position and flux, their formal uncertainties are underestimated. Therefore we want to look into other methods to obtain realistic uncertainties.

The Galactic Centre (GC) is a good example to study crowded fields. Obtaining sensitive, high-angular resolution photometry and astrometry is mandatory to study it, due its unique observational challenges, such as high extinction and extreme stellar crowding. We are interested specially in the inner region around the super massive black hole, Sagittarius \,A* (Sgr\,A*). This region is the most crowded region in the GC, therefore it is imperative to push the angular resolution to the limit and to estimate realistic uncertainties to obtain accurate photometry and astrometry.

There are many different methods for error estimation. Here, we study the bootstrapping \cite{andrae2010error,efron1979bootstrap} resampling method. The bootstrapping method uses "N" measurements (in our case "N" frames in one epoch) in order to create different data sets (in our case, different images). Moreover, the bootstrap samples are produced with replacement, which means that the same data point, can occur multiple times in our bootstrap sample. The main advantage is that the bootstrap method assume that the measured data sample itself contains the information about its error distribution.

We compare three methods for uncertainty estimation. In the first method we apply the {\it StarFinder} (SF) code to a deep image with the full data set and use the formal uncertainties from the program. In the second method we separate the data into three independent sets with 1/3 of the frames each, create three images and analyse them with SF, obtaining the uncertainties from the error of the mean of each star. Method three consists of producing 100 deep images via bootstrapping, with the uncertainties derived from the distribution of the StarFinder measurements on each of them.
We compare the different methods and study their advantages and disadvantages. With the purpose of building a generic procedure to deal with different data sets, we apply the three methods to AO and speckle data, respectively. This article aims at addressing a process to get robust and reliable uncertainties.

\section{OBSERVATIONS AND DATA REDUCTION}

\begin{figure} [ht]
   \begin{center}
   \begin{tabular}{c} %% tabular useful for creating an array of images 
   \includegraphics[width=\textwidth]{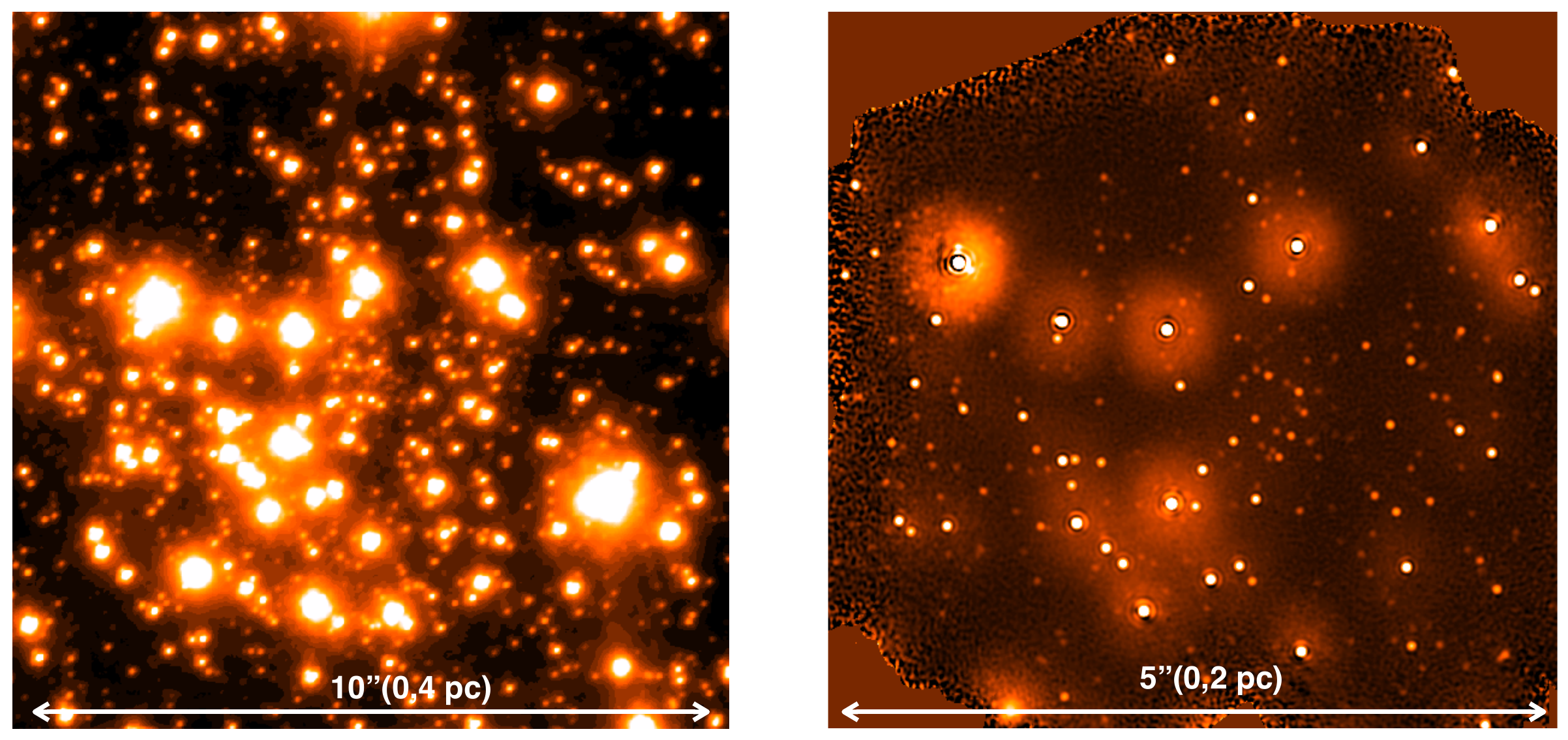}
  \end{tabular}
   \end{center}
   \caption[Figure1] 
%>>>> use \label inside caption to get Fig. number with \ref{}
   { \label{fig:Figure1} 
Imaging data used in the analysis. Left: AO SSA image from 9th August 2008 obtained with NACO/VLT. The field-of-view is $10"\times10"$ centered in Sg\,A*. Right: Speckle holography image from 23th May 2002 obtained with NIRC/Keck. The field-of-view is $5"\times5"$ centered in Sg\,A*.}
   \end{figure}

On the one hand, we use $Ks$-band data from 08-09-2012 obtained with the S27 camera ($0.027"$ pixel scale) on NACO/VLT. The AO was locked on the NIR (near-infrared) bright supergiant GCIRS\,7 that is located about $5.5"$ north of Sgr\,A*. The details of the observation are: $\lambda_{\rm central}$ = 2.18 $\mu$m, $\Delta\lambda$ = 0.35 $\mu$m, the number of dithered expose (N) is 8, the number of integrations that were averaged on-line by the read-out electronics (NDIT) is 60 and the detected integral time (DIT) is  one second. The total integration time of each observation amounts to 480 seconds (N$\times$NDIT$\times$DIT). We want to analyse the most crowded region around Sagittarius A*. For this purpose, we consider the inner part with a field of view of $10"\times10"$ centered in the black hole (See Fig.~\ref{fig:Figure1}, on the left). We have 1920 frames for the epoch.
   
The data and their reduction are presented in Gallego-Cano et al. (2017) \cite{gallego2018distribution}. The main difference to the cited work is that here we create a large number of bootstrap samples to which we then apply the simple shift-and-add (SSA) procedure to obtain final images, as we see in section $3$. 

On the other hand, we use $K$-band data from 04-23-2002 obtained with NIRC ($0.02"$ pixel scale and $\lambda_{\rm central}$ = 2.2 $\mu$m) on the W. M. Keck I telescope. During the epoch of observation, $13440$ frames were obtained using very short exposure times ($0.1$ s) to freeze the distorting effects of the Earth’s atmosphere. We apply a new improved version of holographic reconstruction technique \cite{schodel2013holographic,schodel2012holographic} to obtain a higher-quality, deeper final image. The field of view in this case is $5"\times5"$ centered in the black hole(See Fig.~\ref{fig:Figure1}, on the right). We have 11834 frames for the epoch.

The reduction of the speckle data is explained in detail in different papers\cite{boehle2016improved,meyer2012shortest,schodel2013holographic}. Our version of the speckle holography technique is based on iteratively improved extraction of the instantaneous PSF from speckle frames and on the optional simultaneous use of
multiple reference stars. The algorithm has been specifically developed for crowded fields. In this case, we use an improved version where we use of multiple reference stars to extract the instantaneous PSF from the individual speckle frames, improve the alignment of the speckle frames before the holo procedure, use a quadratic interpolation with a rebinning factor of two, estimate and subtract of a variable sky from each frames and improve the secondary star subtraction of a variable sky from each frame. We obtain a higher-quality, deeper final image.

\section{ASTROMETRY AND PHOTOMETRY}

\begin{figure} [ht]
   \begin{center}
  \begin{tabular}{c} %% tabular useful for creating an array of images 
   \includegraphics[width=\textwidth]{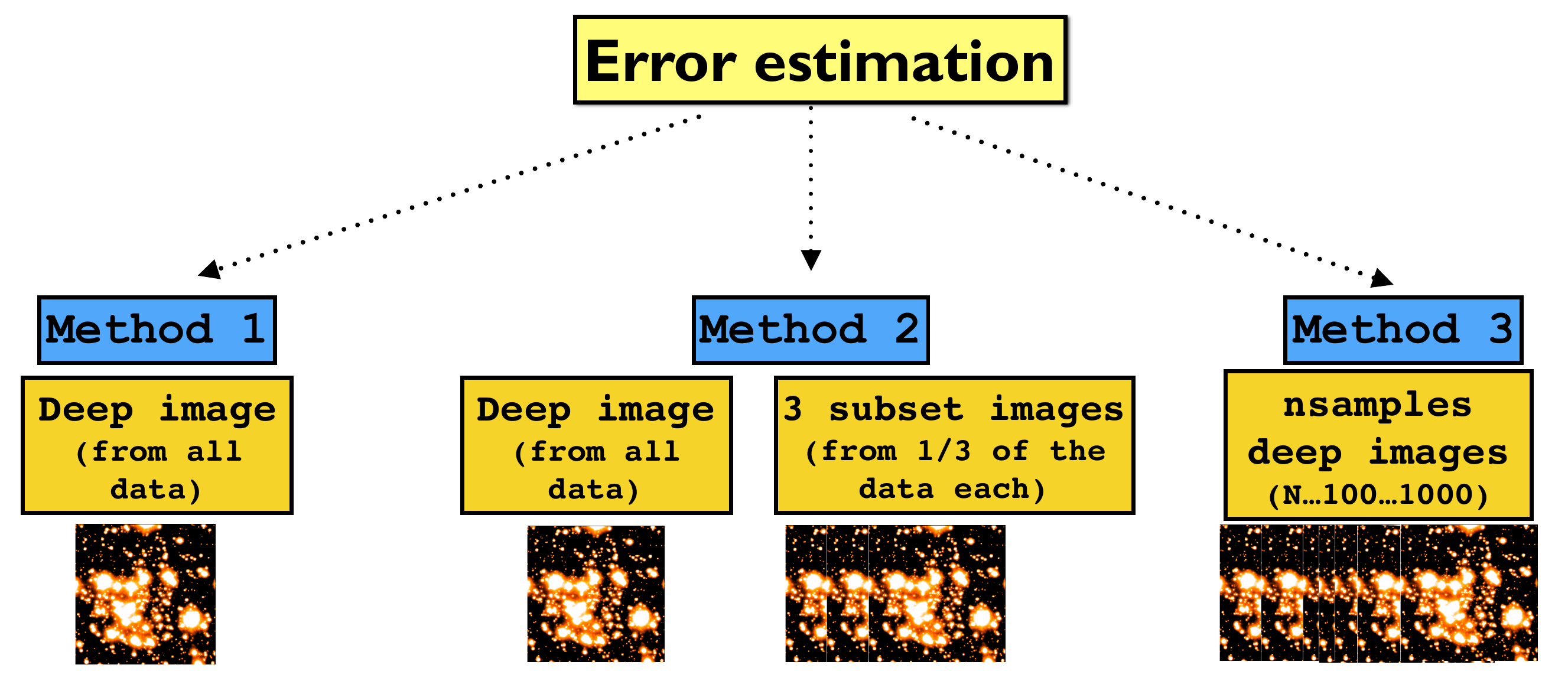}
   \end{tabular}
   \end{center}
   \caption[Figure2] 
%>>>> use \label inside caption to get Fig. number with \ref{}
   { \label{fig:Figure2} 
Scheme of the different methods that we used to obtain the photometric and astrometric uncertainties. For AO, in method 1 we obtained a co-added image with the full data set, run {\it StarFinder} on the image and used the formal uncertainties from  {\it StarFinder}. In method 2 we created three separate data sets with 1/3 of the frames each, created co-added images and analysed them with {\it StarFinder}. Finally we obtained realistic uncertainties from independent data obtained from the error of the mean of each star. In method 3 we created 100 bootstrap samples and corresponding co-added images, run SF on each bootstrapped images and computed the uncertainties from standard deviation of measurements of each star. For speckle data the procedure is the same, but we created co-added images not from samples of individual frames, but from samples of holographically reduced batches of speckle frames.}
   \end{figure}

In this section we explain the different methods that we tested to obtain the astrometry and photometry of the detected stars as well as the associated astrometric and photometric uncertainties. Fig.~\ref{fig:Figure2} shows an outline of the procedure that we followed
after the basic reduction for each method.

\subsection{AO data}

\subsubsection{Method 1}
\begin{figure} [ht]
   \begin{center}
  \begin{tabular}{c} %% tabular useful for creating an array of images 
   \includegraphics[width=\textwidth]{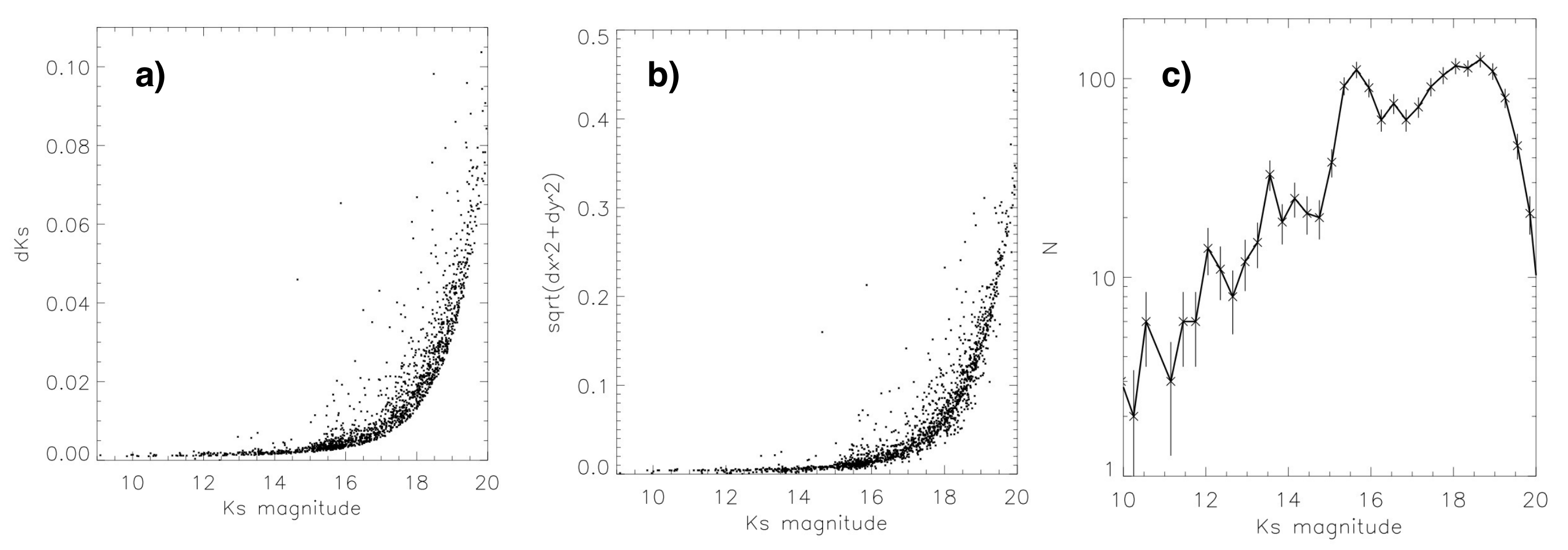}
   \end{tabular}
   \end{center}
   \caption[Figure3] 
%>>>> use \label inside caption to get Fig. number with \ref{}
   { \label{fig:Figure3} 
Results obtained applying method $1$ to AO data. a) Photometric uncertainties versus $K_{s}$ magnitude from {\it Starfinder} program. b) Astrometric uncertainties versus $K_{s}$ magnitude from {\it Starfinder} program. We represent $sqrt(dx^{2}+dy^{2})$ versus $K_{s}$, where $dx$ and $dy$ are the errors in the x-position and y-position, respectively, of each star. c) KLF for the deep image.}
   \end{figure}

First, a simple SSA procedure was applied to obtain the final deep image considering all the frames in the epoch. After that, we used the SF program to detect and subtract detected point sources from the image. We are interested in studying the resulting uncertainties of real stars and we want to avoid the possible contamination of spurious sources, hence we used the following conservative values for the {\it StarFinder} parameters:  {\it min\_correlation$=0.80$} and {\it deblend$ = 0$}. The detection threshold was chosen as $3\sigma$ and we applied two iteration of SF algorithm. The photometry was calibrated with the stars IRS\,16C,IRS\,16NW, and IRS\,33N (apparent magnitudes $K_{s}=9.93,10.14,11.20$ see\cite{schodel2010peering}). In Fig.~\ref{fig:Figure3} we show the photometric and astrometric uncertainties obtained along with the Ks-luminosity function (KLF). We detected $1618$ stars and the detection limit, that we define as the value of the magnitude where the cumulative number of all detected stars reached $90$\%, is $19.97$. The median value of the photometric uncertainty for brighter stars ($K_{s} < 14$)is $0.002$.

\subsubsection{Method 2}

\begin{figure} [ht]
   \begin{center}
  \begin{tabular}{c} %% tabular useful for creating an array of images 
   \includegraphics[width=\textwidth]{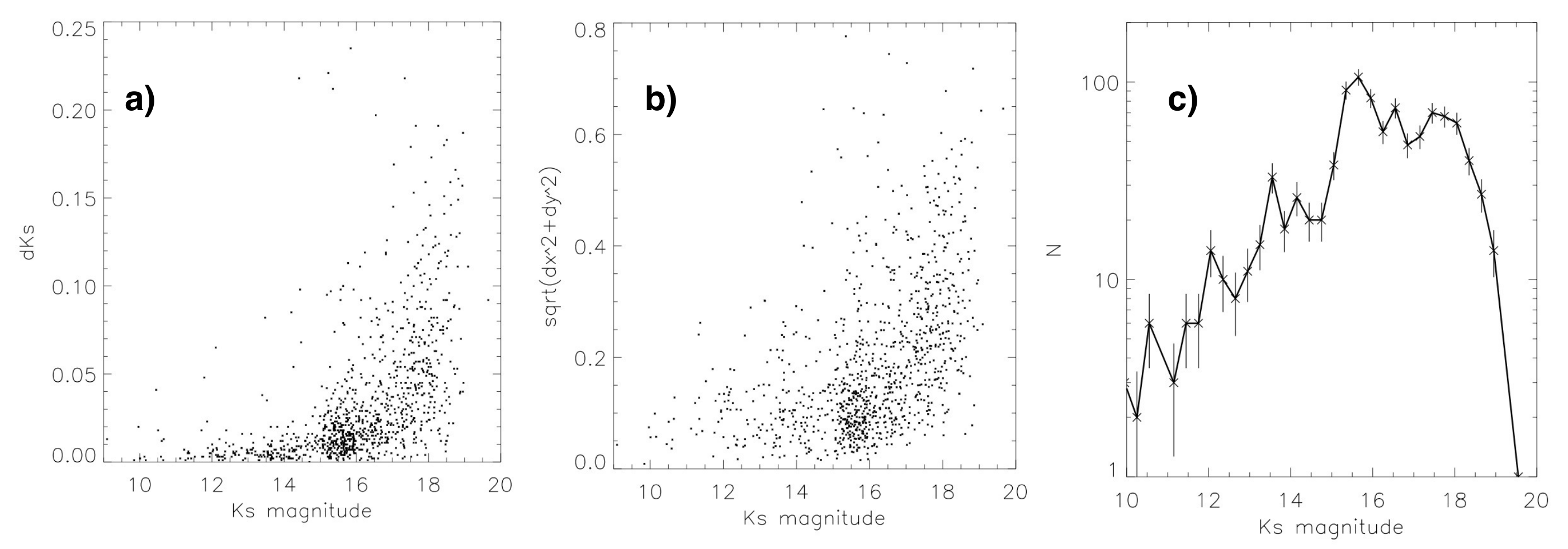}
   \end{tabular}
   \end{center}
   \caption[Figure4] 
%>>>> use \label inside caption to get Fig. number with \ref{}
   { \label{fig:Figure4} 
Results obtained applying method $2$ to AO data. a) Photometric uncertainties versus $K_{s}$ magnitude from the errors of the mean of each star detected in all images. b) Astrometric uncertainties versus $K_{s}$ magnitude c) KLF for the stars common to all images.}
   \end{figure}

First, we divided the full data in 3 separate data sets with $1/3$ of frames each. A simple SSA procedure was applied to each subset to obtain three final images. After that, we used the {\it StarFinder} program to detect and subtract detected point sources from the three images. We used the same values for the {\it StarFinder} parameters than for method 1. We compared detected stars in the three subset and considered the common stars. We calibrated the images similar to method $1$, but in this case we corrected the offset in the positions and zero points that we could have due to small differences between their estimated PSFs. Finally, we computed uncertainties from the error of the mean of the position and flux of each star. In Fig.~\ref{fig:Figure4} we show the photometric and astrometric uncertainties obtained from {\it Starfinder} and the KLF.  We detected $1032$ common stars and the detection limit is $18$. 
The median value of the photometric uncertainty for brighter stars with $K_{s} < 14$ is $0.005$.

\subsubsection{Method 3}

\begin{figure} [ht]
   \begin{center}
  \begin{tabular}{c} %% tabular useful for creating an array of images 
   \includegraphics[width=\textwidth]{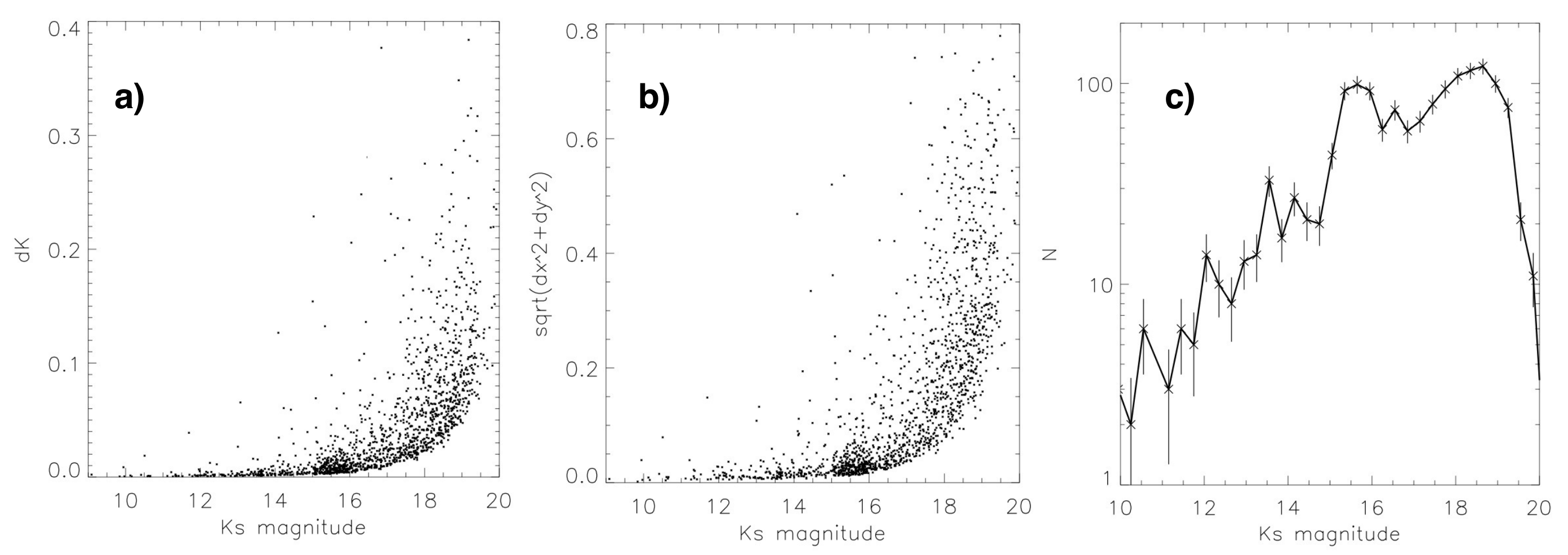}
   \end{tabular}
   \end{center}
   \caption[Figure5] 
%>>>> use \label inside caption to get Fig. number with \ref{}
   { \label{fig:Figure5} 
Results obtained applying method $3$ to AO data for  {\it detection frequency}=50\%. a) Photometric uncertainties versus $K_{s}$ magnitude from the standard deviation of the measurements of each star. b) Astrometric uncertainties versus $K_{s}$ magnitude. c) KLF for the final list.} 
   \end{figure}

First, we created $100$ SSA images using the bootstrapping with replacement method. The number of frames considered in each bootstrap image is equal to the total number of frames. After that, we used the SF program to detect and subtract detected point sources from $100$ bootstrap images. We used the same values for the SF parameters than for the previous methods. We defined the {\it detection frequency} parameter as the percentage of bootstrap deep images where each star is detected and we obtained a final list with all the stars detected in all the images and their associated detection frequency value. We selected {\it detection frequency}=50\%, which means that we considered stars detected in 50\% or more of the bootstrap images. We calibrated the images similar to method $2$, correcting the offset in the positions and zero points that we could have between all the images, too. Finally, we computed uncertainties from the standard deviation of the position and flux of each star. In Fig.~\ref{fig:Figure5} we show the photometric and astrometric uncertainties obtained with this method and the KLF.  We detected $1514$ stars and the detection limit is $19$. 
The median value of the photometric uncertainty for brighter stars with $K_{s} < 14$ is $0.002$.

\subsection{Speckle data}
The procedure that we followed is the same than for AO data. The only difference is that in this case we first run holography on independent subsamples of the speckle frames and we combined the resulting  holographic images, as we see in Section 2.

\subsubsection{Method 1}
\begin{figure} [ht]
   \begin{center}
  \begin{tabular}{c} %% tabular useful for creating an array of images 
   \includegraphics[width=\textwidth]{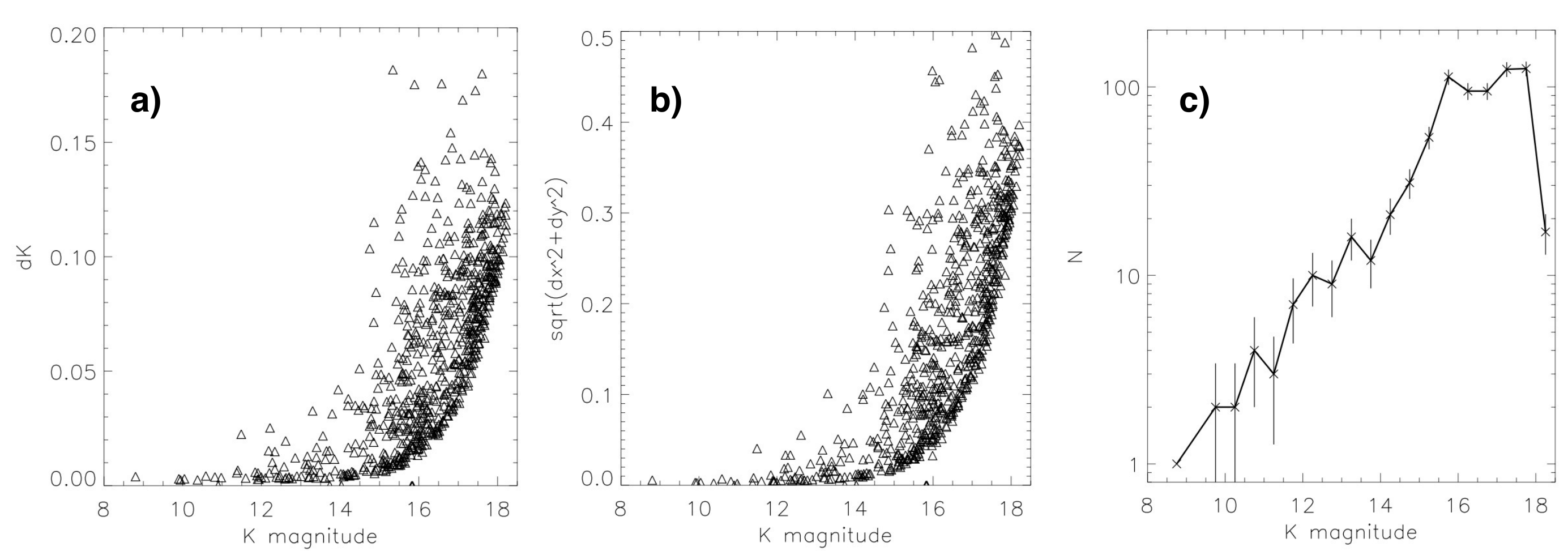}
   \end{tabular}
   \end{center}
   \caption[Figure_spm1] 
%>>>> use \label inside caption to get Fig. number with \ref{}
   { \label{fig:Figure_spm1} 
Results obtained applying method $1$ to speckle data. a) Photometric uncertainties versus $K$ magnitude from {\it Starfinder} program. b) Astrometric uncertainties versus $K$ magnitude from {\it Starfinder} program. c) KLF for the deep image.}
   \end{figure}

In this case, due to the small field of NIRC, we study the central $5"$ around Sgr\,A*. We used the following values for the SF parameters:  {\it min\_correlation$=0.80$} and {\it deblend$ = 1$}. The other SF parameters are the same than for the AO data. The photometry was calibrated with the stars IRS\,16C, IRS\,16NW, and IRS\,16CC (apparent magnitudes $K=9.83,10.03,10.36$ see\cite{blum1996jhkl}). In Fig.~\ref{fig:Figure_spm1} we show the photometric and astrometric uncertainties obtained and the K-luminosity function (KLF). We detected $741$ stars and the detection limit is $18$. The median value of the photometric uncertainty for brighter stars ($K_{s} < 14$) is $0.005$.

\subsubsection{Method 2}

\begin{figure} [ht]
   \begin{center}
  \begin{tabular}{c} %% tabular useful for creating an array of images 
   \includegraphics[width=\textwidth]{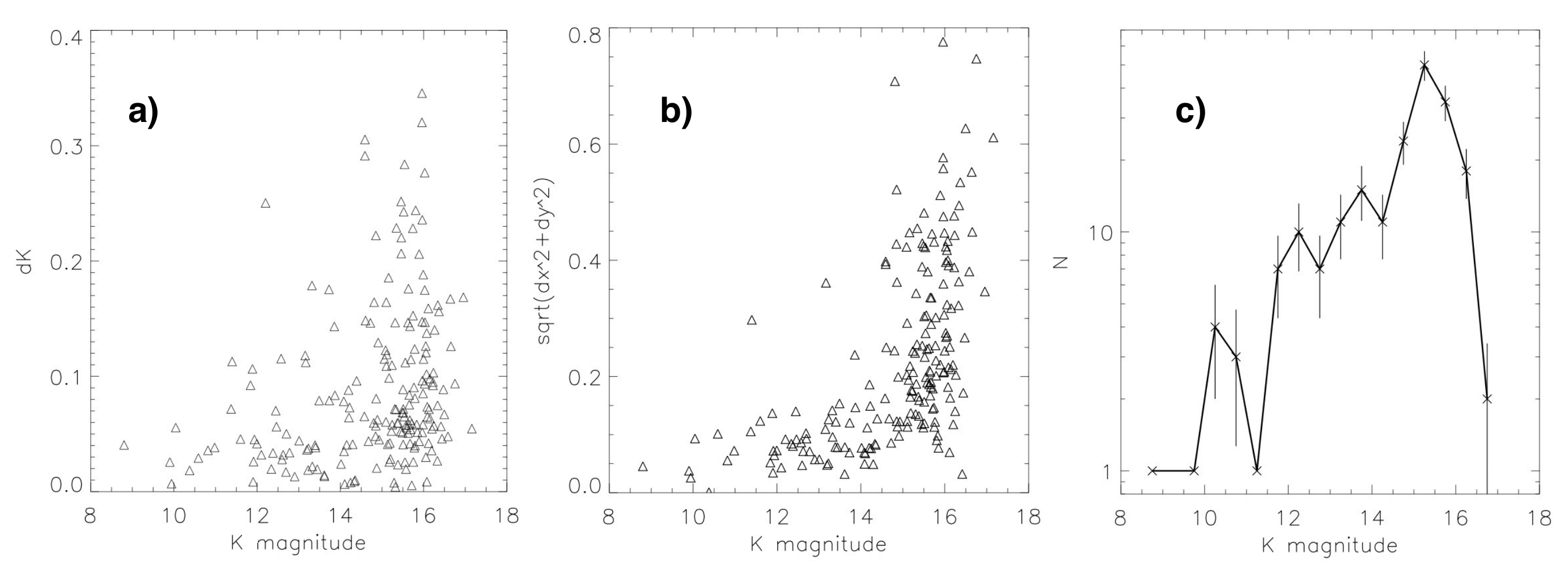}
   \end{tabular}
   \end{center}
   \caption[Figure_spm2] 
%>>>> use \label inside caption to get Fig. number with \ref{}
   { \label{fig:Figure_spm2} 
Results obtained applying method $2$ to speckle data. a) Photometric uncertainties versus $K$ magnitude from the errors of the mean of each star detected in all images. b) Astrometric uncertainties versus $K$ magnitude. c) KLF for the common stars to all images.}
   \end{figure}

After the holographic procedure, we obtained $76$ holo images. The procedure is the same than for AO data (see section $3.1.2$). We used the same values for the {\it StarFinder} parameters than for Method $1$. Finally, we computed uncertainties from the error of the mean of the position and flux of each star. In Fig.~\ref{fig:Figure_spm2} we show the photometric and astrometric uncertainties obtained from {\it Starfinder} and the KLF.  We detected $200$ common stars and the detection limit is $16$. 
The median value of the photometric uncertainty for brighter stars with $K_{s} < 14$ is $0.06$.

\subsubsection{Method 3}

\begin{figure} [ht]
   \begin{center}
  \begin{tabular}{c} %% tabular useful for creating an array of images 
   \includegraphics[width=\textwidth]{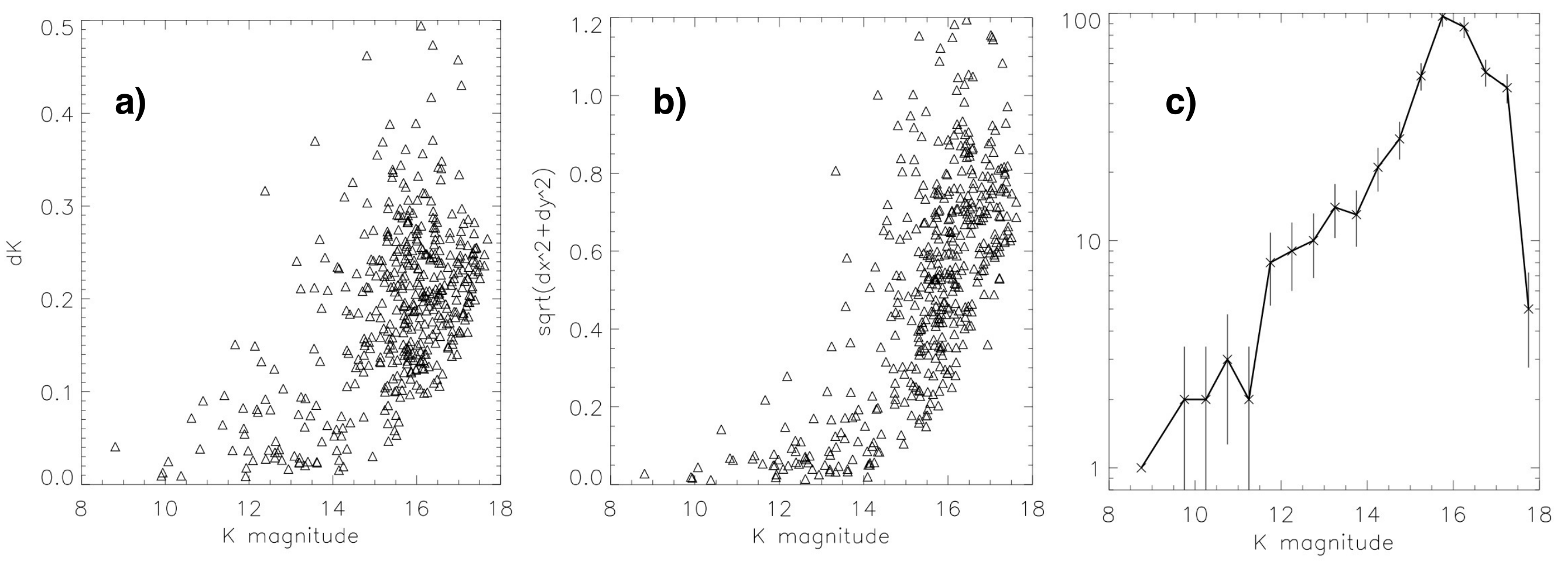}
   \end{tabular}
   \end{center}
   \caption[Figure_spm3] 
%>>>> use \label inside caption to get Fig. number with \ref{}
   { \label{fig:Figure_spm3} 
Results obtained applying method $3$ to speckle data for  {\it detection frequency}=50\%. a) Photometric uncertainties versus $K$ magnitude from the standard deviation of the measurements of each star. b) Astrometric uncertainties versus $K$ magnitude. c) KLF for the final list.} 
   \end{figure}

In this case, we carry out a bootstrapping resampling method similar to AO (see section $3.1.3$). The only difference is that now we bootstrap holographically pre-reduced images and not bootstrap individual frames, as in the AO case, with the aim of speed up the process. We use the same values for the {\it StarFinder} parameters that for Method $1$ and $2$. We select {\it detection frequency}=50\%. In Fig.~\ref{fig:Figure_spm3} we show the photometric and astrometric uncertainties obtained with this method and the KLF.  We detect $457$ stars and the detection limit is $17.05$. 
The median value of the photometric uncertainty for brighter stars with $K_{s} < 14$ is $0.06$.

\section{DISCUSSION}

\begin{figure} [ht]
   \begin{center}
  \begin{tabular}{c} %% tabular useful for creating an array of images 
   \includegraphics[width=\textwidth]{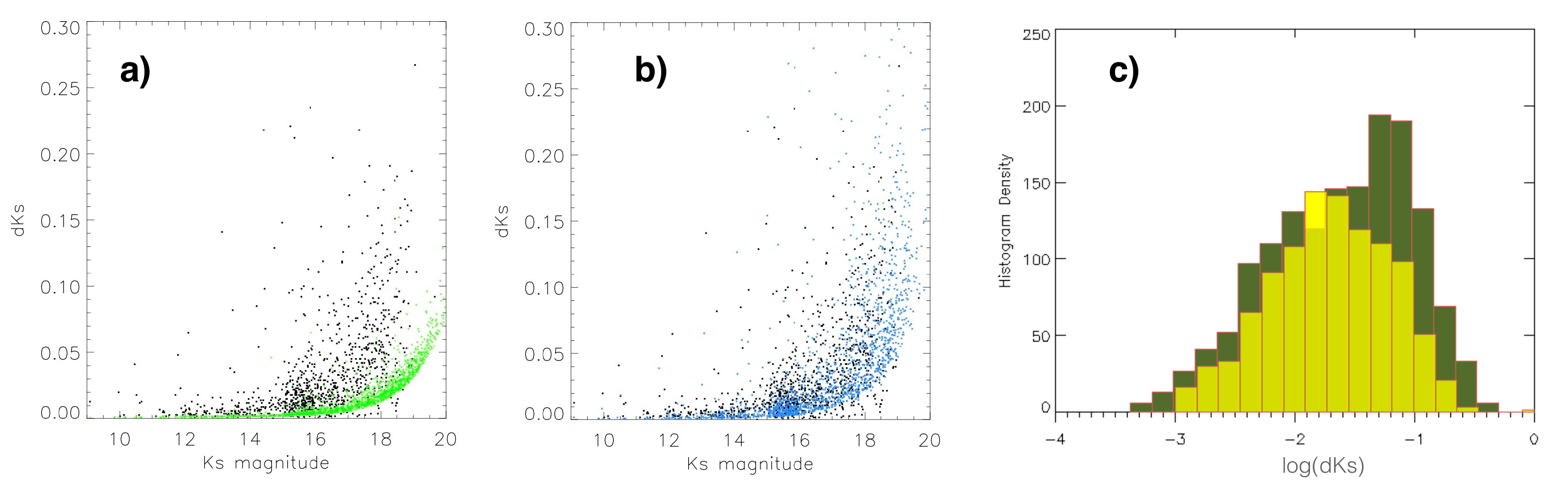}
   \end{tabular}
   \end{center}
   \caption[Figure6] 
%>>>> use \label inside caption to get Fig. number with \ref{}
   { \label{fig:Figure6} 
Final comparison between the three methods for AO. a) Comparison between the photometric uncertainties obtained by method $1$ (green) and method $2$ (black). b) Comparison between the photometric uncertainties obtained by method $2$ (black) and method $3$ (blue).c) Distributions of errors obtained by method $2$ (yellow) and  method $3$ (green).} 
   \end{figure}

\begin{figure} [H]
   \begin{center}
  \begin{tabular}{c} %% tabular useful for creating an array of images 
   \includegraphics[width=\textwidth]{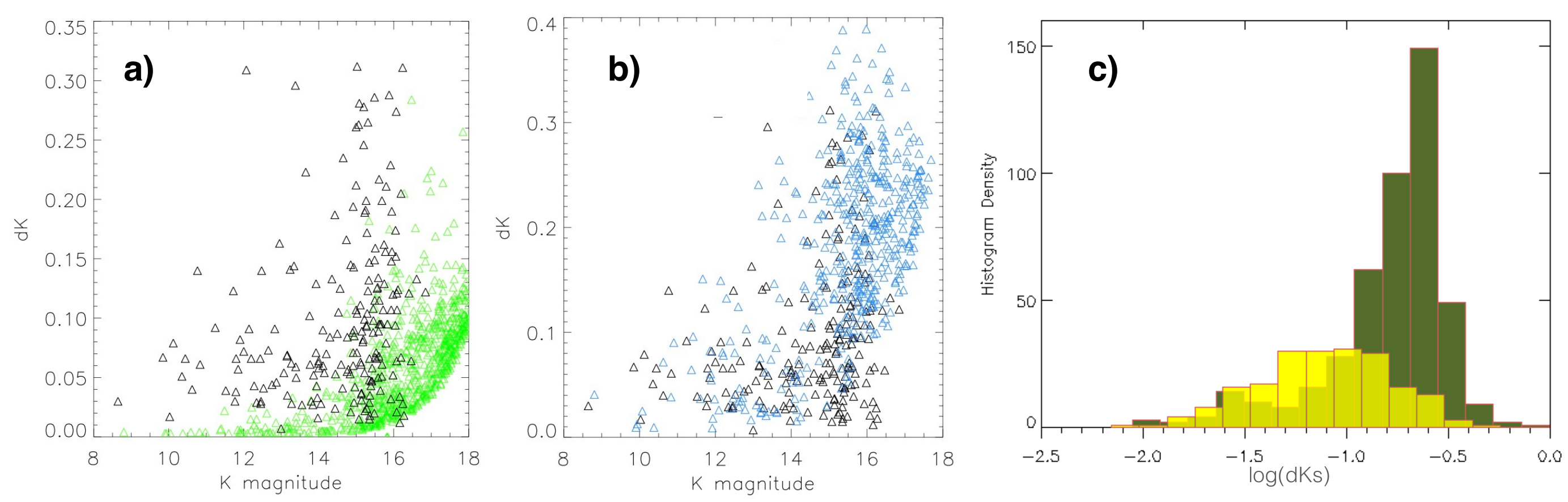}
   \end{tabular}
   \end{center}
   \caption[resultspeck] 
%>>>> use \label inside caption to get Fig. number with \ref{}
   { \label{fig:resultspeck} 
Final comparison between the three methods for speckle. a) Comparison between the photometric uncertainties obtained by method $1$ (green) and method $2$ (black). b) Comparison between the photometric uncertainties obtained by method $2$ (black) and method $3$ (blue).c) Distributions of errors obtained by method $2$ (yellow) and  method $3$ (green).} 
   \end{figure}

In this section, we compare the results obtained by the three methods. Fig.~\ref{fig:Figure6} and Fig.~\ref{fig:resultspeck} show the comparison between the different methods for AO data and speckle data, respectively. If we compare the uncertainties obtained by method $2$ with the uncertainties obtained by method $1$, we can see that {\it StarFinder} under-estimates the uncertainties (see $a$ in Fig.~\ref{fig:Figure6} and Fig.~\ref{fig:resultspeck}). If we compare the results obtained by the method $2$ and method $3$, we can see that we obtain similar uncertainties (see $b$ in Fig.~\ref{fig:Figure6} and Fig.~\ref{fig:resultspeck}). Therefore, we can get statistically similar uncertainties by both method $2$ and $3$. 

\begin{figure} [H]
   \begin{center}
   \begin{tabular}{c} %% tabular useful for creating an array of images 
   \includegraphics[width=\textwidth]{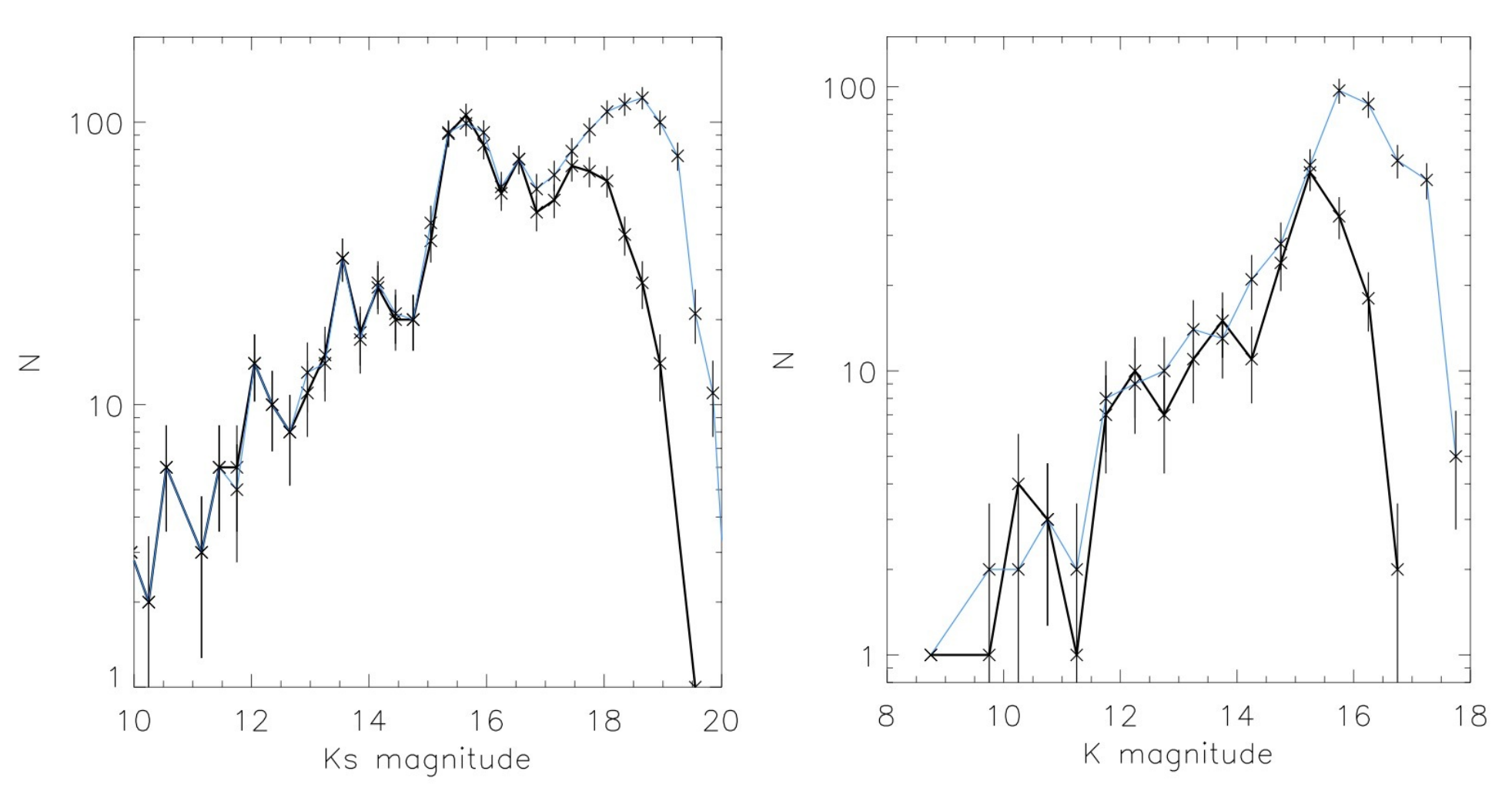}
  \end{tabular}
   \end{center}
  \caption[Figure7] 
%>>>> use \label inside caption to get Fig. number with \ref{}
   { \label{fig:Figure7} 
Comparison between KLFs obtained by method $2$ and $3$ for AO data(on the left) and for speckle data(on the right). Both give robust uncertainty estimates, but method $3$ is more sensitive.}
 \end{figure}

The Fig.~\ref{fig:Figure7}) compare KLFs obtained by method $2$ and $3$, respectively. We can see that although both methods give reliable uncertainties, we can go almost one magnitude deeper by applying the bootstrapping procedure than by applying method $2$.
   
\section{CONCLUSIONS}
In this article, we show three methods to obtain uncertainties. We see that {\it StarFinder} under-estimates the uncertainties. We obtain robust uncertainty estimates by separating the data into disjunct subsets, but at the cost of loss of sensitivity. We show a new method that uses bootstrapping resampling approach. This method allows us to obtain robust estimates of the uncertainty and is more sensitive than method 2.

\acknowledgments % equivalent to \section*{ACKNOWLEDGMENTS}       
The research leading to these results has received funding from the European Research Council under the European Union's Seventh Framework Programme (FP7/2007-2013) / ERC grant agreement n$^{\circ}$ [614922]. This work is based on observations made with ESO Telescopes at the La Silla Paranal Observatory under programm ID089.B-0162.

% References
\bibliography{report.bib} % bibliography data in report.bib
%\bibliography{/Users/Laly/BibDesk/BibGC}

\bibliographystyle{spiebib} % makes bibtex use spiebib.bst

\end{document}